\documentclass{ws-procs975x65}

\usepackage[pdftex,breaklinks=false,bookmarksnumbered]{hyperref}

\def\nl{\\ & \quad}

\allowdisplaybreaks

\begin{document}

\title{ADM canonical formulation with spin\\and application to post-Newtonian approximations}

\author{J.~Steinhoff, S.~Hergt, and G.~Sch\"afer}

\address{Theoretisch-Physikalisches Institut, Friedrich-Schiller-Universit\"at,\\
Max-Wien-Platz\ 1, 07743 Jena, Germany, EU}

\begin{abstract}
Recently, different methods succeeded in calculating the spin dynamics at higher orders in the post-Newtonian (PN) approximation. This is an essential step toward the determination of more accurate templates for gravitational waves, to be used in future gravitational wave astronomy. We focus on the extension of the ADM canonical formalism to spinning binary black holes. Using the global Poincar\'e invariance of asymptotically flat spacetimes as the most important guiding consistency condition, this extension can be constructed order by order in the PN approximation. We were able to reach a high order both in the spin power and the PN counting.
\end{abstract}


\bodymatter

\section{ADM Hamiltonian and global Poincar\'e invariance}
In this article we consider asymptotically flat spacetimes and make use of
standard (3+1)-splitting of spacetime into a stack of 3-dim.\ spacelike hypersurfaces.
The Einstein equations are split into constraints and evolution equations.
The speed of light $c$ and the gravitational constant $G$ are put equal to one.

The ADM Hamiltonian\cite{Arnowitt:Deser:Misner:1962} $H[\hat{z}^i_a, P_{ai}, \hat{S}_{a(i)(j)}, h^{\text{TT}}_{ij}, \pi^{ij\text{TT}}_{\text{can}}]$ basically
is the ADM energy expressed in terms of variables which possess standard canonical Poisson brackets
(called canonical variables in the following).
Here the matter variables $\hat{z}^i_a$, $P_{ai}$, and $\hat{S}_{a(i)(j)}$ are the canonical
position, momentum, and spin tensor of the $a$-th object, respectively. $h^{\text{TT}}_{ij}$ is
the transverse-traceless part of the induced metric of the spacelike hypersurfaces and
$\pi^{ij\text{TT}}_{\text{can}}$ its canonical conjugate momentum (related to the extrinsic curvature).
In order to calculate the ADM Hamiltonian, one only has to solve the field constraints
(in the ADM gauge), not the evolution equations.

The global Poincar\'e algebra is a consequence of the
asymptotic flatness and is represented by Poisson brackets
of the corresponding conserved quantities\cite{Regge:Teitelboim:1974}. These quantities
are the ADM energy $H$, total linear momentum $P_i$,
total angular momentum tensor $J_{ij}$,
and the boost vector $K_i$. The boosts have an explicit dependence on the time
$t$ and can be decomposed as $K_i = G_i - t P_i$.

\section{Hamiltonians linear in the single spin variables}
We treat black holes not as vacuum solutions of the Einstein equations. Instead we
represent them by Tulczyjew's singular stress-energy tensor\cite{Tulczyjew:1959} $T_{\mu\nu}$
(in the covariant spin supplementary condition), which contains the matter variables.
The matter variables are evolved by the Mathisson-Papapetrou equations\cite{Mathisson:1937}.
Then it is straightforward to calculate the ADM energy or the other conserved quantities
depending on the matter variables appearing in $T_{\mu\nu}$, at least up to some order in the
post-Newtonian (PN) approximation (using certain regularization
techniques). However, the variables appearing in $T_{\mu\nu}$
are, in general, not standard canonical and one has to redefine them. One approach is to find such
a redefinition by requiring that the conserved quantities fulfill the
global Poincar\'e algebra. Then it must hold\cite{Steinhoff:Wang:2009}
\begin{subequations}\label{PJ}
\begin{align}
	P_i &= \sum_a P_{ai} - \frac{1}{16\pi} \int d^3x \, \pi_{\text{can}}^{kl\text{TT}} h^{\text{TT}}_{kl,i} \,, \label{PJ1} \\
\begin{split}\label{PJ2}
	J_{ij} &= \sum_a ( \hat{z}_a^i p_{aj} - \hat{z}_a^j p_{ai} ) + \sum_a \hat{S}_{a(i)(j)}
		- \frac{1}{16\pi} \int d^3x \, ( x^i \pi_{\text{can}}^{kl\text{TT}} h^{\text{TT}}_{kl,j} \nl - x^j \pi_{\text{can}}^{kl\text{TT}} h^{\text{TT}}_{kl,i} )
		- \frac{1}{8\pi} \int d^3x \, ( \pi_{\text{can}}^{ik\text{TT}} h^{\text{TT}}_{kj} - \pi_{\text{can}}^{jk\text{TT}} h^{\text{TT}}_{ki} ) \,,
\end{split}
\end{align}
\end{subequations}
which implies that a major part of the Poincar\'e algebra is fulfilled. Requiring that $P_i$
and $J_{ij}$ take this form indeed fixes the transition to canonical variables in a unique way
at the next-to-leading order (NLO)\cite{Steinhoff:Schafer:Hergt:2008,Steinhoff:Wang:2009}.
This allowed us to compute the NLO spin-orbit Hamiltonian\cite{Steinhoff:Schafer:Hergt:2008},
which has already been derived earlier\cite{Damour:Jaranowski:Schafer:2008:1} and agrees with
non-canonical results\cite{Tagoshi:Ohashi:Owen:2001,Faye:Blanchet:Buonanno:2006}.
A new result was the complete NLO S$_1$S$_2$ Hamiltonian\cite{Steinhoff:Hergt:Schafer:2008:1},
which was later confirmed by a different method\cite{Porto:Rothstein:2008:1}. Further,
$G_i$ was calculated and the full Poincar\'e algebra was checked\cite{Steinhoff:Hergt:Schafer:2008:1,Steinhoff:Schafer:Hergt:2008}.

It was shown that at higher PN orders one also has to redefine the canonical
momentum of the gravitational field\cite{Steinhoff:Wang:2009}. By considering
Eqs.\ (\ref{PJ}), all variable redefinitions necessary for the next-to-next-to-leading order
can be found (up to a canonical transformation)\cite{Steinhoff:Wang:2009}.
Variable redefinitions leading to canonical variables valid at all orders were recently
constructed using an action approach\cite{Steinhoff:Schafer:2009:2}.
The used action basically results from a minimal coupling of the flat space one\cite{Hanson:Regge:1974},
similar as in previous approaches\cite{Porto:2006}. Next, all
constraints, supplementary, gauge, and coordinate conditions are eliminated from
the action. The variable redefinitions in question then transform this action
into the form known to produce Hamilton's equations.

\section{Hamiltonians non-linear in the single spin variables}
If one knows the singular source terms of the constraints in terms of canonical
variables, a straightforwardly calculated ADM energy will automatically be the sought-for Hamiltonian.
This approach was used for higher orders in the spin for binary black holes.
Source terms in canonical variables sufficient for $H_{S_2^2S_1p_1}$, $H_{S_2^3p_1}$, $H_{S_1^3p_2}$, $H_{S_1^2S_2p_2}$, $H_{S_1^2S_2^2}$, $H_{S_1S_2^3}$, and $H_{S_2S_1^3}$ were obtained from the Kerr metric in ADM coordinates\cite{Hergt:Schafer:2008:2}.
All these Hamiltonians are linear in $G$.
The relation between Kerr parameter and $J_{ij}$ is crucial to obtain the canonical spin.
Further, a leading order boost was applied to the Kerr metric, which allowed for leading corrections in the momenta.

Another approach is to construct $H$ and $G_i$ depending on canonical variables
directly from an ansatz and using the global Poincar\'e algebra to fix the
coefficients, with $P_i$ and $J_{ij}$ still given by Eqs.\ (\ref{PJ}).
Ansatzes for $H_{S_{1}^2p^2}$, $H_{S_{2}^2p^2}$, $H_{S_{1}^3p_{1}}$, $H_{S_{2}^3p_{2}}$, $H_{S_{1}^2S_{2}p_{1}}$, $H_{S_{2}^2S_{1}p_{2}}$, $H_{S_{1}^4}$, and $H_{S_{2}^4}$ (again all linear in $G$) can indeed
be fixed up to canonical transformation by $\left\{G_{i},H\right\}=P_{i}$ only\cite{Hergt:Schafer:2008}.
Here also a corresponding ansatz for the source terms was used.
The test mass case was used as a check.

The static ($P_{ai}$-independent) source part of the Hamilton constraint at NLO S$_1^2$ is needed to fix the remaining degrees of freedom from the canonical transformation,
as well as to get the missing $G^2$-part of the NLO S$_1^2$ Hamiltonian
(the linear-in-$G$ part is given by $H_{S_{1}^2p^2}$). A general ansatz for these source terms,
however, contains only four unknown constants\cite{Steinhoff:Hergt:Schafer:2008:2}. Two of these constants could be fixed by matching to the Kerr metric in ADM coordinates.
The remaining two constants do not contribute to the Hamiltonian due to some cancellations.
Finally, the complete NLO S$_1^2$ Hamiltonian can be calculated\cite{Steinhoff:Hergt:Schafer:2008:2}.
The spin evolution agrees with an earlier result\cite{Porto:Rothstein:2008:2,Steinhoff:Schafer:2009:1}
and the test mass case was checked.
If one wants to use the approach of the last section to derive the NLO S$_1^2$ Hamiltonian,
one needs quadrupole corrections to Tulczyjew's stress-energy tensor
\cite{Steinhoff:Puetzfeld:2009}. This, however, has not been tried yet.

It should be noted that all Hamiltonians for binary black holes at the formal 2PN order are now known
(for the formal counting, see, e.g., Ref.\ \refcite{Hergt:Schafer:2008} and also Appendix A of Ref.\ \refcite{Steinhoff:Wang:2009}).

\section*{Acknowledgments}
This work is supported by the Deutsche Forschungsgemeinschaft (DFG) through
SFB/TR7 ``Gravitational Wave Astronomy''.

\providecommand{\href}[2]{#2}\begingroup\raggedright\endgroup

\end{document}